# A Computational Approach for Mapping Electrochemical Activity of Multi-Principal Element Alloys

Jodie A. Yuwono[1,2], Xinyu Li[3], Tyler D. Doležal[4], Adib J. Samin[4],
Javen Qinfeng Shi[3], Zhipeng Li[2] and Nick Birbilis[2,5]

[1]School of Chemical Engineering, The University of Adelaide, Adelaide, SA 5005, Australia
[2]College of Engineering, Computing and Cybernetics, Australian National University, Canberra, ACT 2601, Australia
[3]Australian Institute for Machine Learning, The University of Adelaide, Adelaide, SA 5000, Australia
[4]Department of Engineering Physics, Air Force Institute of Technology, 2950 Hobson Way, Wright-Patterson Air Force Base, OH 45433, USA
[5]Faculty of Science, Engineering and Built Environment, Deakin University, Waurn Ponds, VIC 3216, Australia

**ABSTRACT**
Multi principal element alloys (MPEAs) comprise a unique class of metal alloys. MPEAs have been demonstrated to possess several exceptional properties, including, as most relevant to the present study a high corrosion resistance. In the context of MPEA design, the vast number of potential alloying elements and the staggering number of elemental combinations favours a computational alloy design approach. In order to computationally assess the prospective corrosion performance of MPEA, an approach was developed in this study. A density functional theory (DFT) – based Monte Carlo method was used for the development of MPEA 'structure'; with the AlCrTiV alloy used as a model. High-throughput DFT calculations were performed to create training datasets for surface activity/selectivity towards different adsorbate species: $O^{2-}$, $Cl^-$ and $H^+$. Machine-learning (ML) with combined representation was then utilised to predict the adsorption and vacancy energies as descriptors for surface activity/selectivity. The capability of the combined computational methods of MC, DFT and ML, as a virtual electrochemical performance simulator for MPEAs was established and may be useful in exploring other MPEAs.

## 1. INTRODUCTION

In conventional alloy design to date, typically a principal metallic element is explored, accompanied by other elements in relatively minor concentrations. Such alloy design – whilst often empirical – has also been aided by thermodynamic calculations; the development and use of phase diagrams; and more recently by kinetic assessment to factor in thermally activated solid-state transformations. Whilst such approaches have developed to significant levels of sophistication, the exploration of more complicated alloy systems, of which no prior data exists, can make desktop alloy design complex[1] Modern alloy exploration in the form of the so-called multi-principal element alloy (MPEAs) requires new tools and approaches to aid in their design. As the name suggests, MPEAs contain at least two (but often more, including >5), principal alloying elements. One impetus for the rapid development and research regarding MPEAs is owing to their excellent physical properties[2] and more recently, reports of their aqueous corrosion resistance – amongst other unique properties. Furthermore, other applications in which MPEAs are being explored include catalysts and battery electrodes, in addition to structural materials[3]. Due to the large number of possible alloying elements available for potential MPEA production, and the myriad of possible elemental combinations, the realm of MPEAs still remains heavily unexplored[4]. If MPEA development is to continue at an appropriate pace, conventional trial-and-error methods for alloy development must be supplemented by objective alloy design[5]. In addition to the need for rational alloy design based on tools that can assist in prediction of structure and perhaps key properties, the desktop computational prediction of corrosion performance remains its own unique challenge. To date, there are very few mechanistic or deterministic tools available that can generate computed information in a manner that can aid in prediction of electrochemical performance. In this work, a computational methodology (which is actually a computational workflow) is presented, in order to estimate the electrochemical performance of an MPEA, without the need for any empirical testing.

Integrated approaches permit computational materials discovery approaches for new (objective oriented) alloy design. To date, the calculation of phase diagrams (CALPHAD), which is based on equilibrium thermodynamics to provide information on probable equilibrium phases, is widely used; as is the use of density functional theory (DFT) to provide information on phase stability. Though CALPHAD and DFT approaches are beneficial, such approaches may be time-consuming, computationally costly and limited in what can be explored relative to properties of interest and the breadth of MPEA compositions[1,6]. Recently, a purely machine learning (ML) approach has been developed to accelerate the discovery of MPEAs, with a specific focus on alloy development of INVAR[7]. This kind of ML-based method provides the capability to build a connection between underlying physics and composition-dependent properties[7,8], allowing for the selection, design and verification of new alloys to be done an order of magnitude faster. However, the lack of understanding regarding how alloy composition can impact material properties for unexplored composition space, remains a bottleneck. This level of understanding is critical for predicting which elemental combinations may yield optimal structural performance in various operational environments.

When a primary property of interest is alloy electrochemistry and/or corrosion behaviour, detailed information regarding surface features is critical – as electrochemical processes occur at the alloy surface (i.e. alloy-environment interface). Without a complete understanding of the composition-dependent surface properties in the context of electrochemistry and/or corrosion, the verification of newly designed alloys will be tedious. For instance, in alloys that develop local sites with different electrochemical activities, this may accelerate corrosion due to micro-galvanic coupling or localised pit formation[9]. Herein, a computational approach is proposed that combines Monte Carlo (MC) methods, high-throughput DFT, and ML – to enable virtual electrochemical characterisation of digitally designed (new) alloys. As a proof of concept, this methodology was explored for an MPEA with the equi-atomic composition AlCrTiV. This alloy was selected as a representative model system, because its electrochemical activity has been thoroughly and well-documented in the literature[10–14]. The 'electrochemistry simulator' is expected to describe the relationship between surface features and surface reactivity. Furthermore, the intent is to provide a pathway, if not a methodology, for data collection and verification;, which would enable a knowledge-guided ML approach for materials discovery of MPEAs with tailored electrochemical properties.

## 2. APPROACH

A state-of-the-art hybrid MC and DFT approach was employed in order to develop MPEA models sampled from equilibrium[15]. Herein, the ability to predict electrochemical activity of an MPEA using computer simulation was investigated. Three stages were employed in this workflow (**Figure 1**).

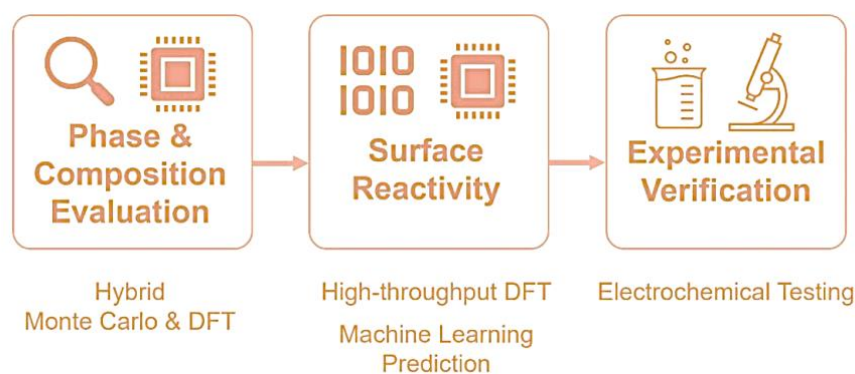

**Figure 1.** *The overall workflow of virtual electrochemistry simulation performed in this study and its verification.*

The phase and composition at equilibrium were obtained using the multi-cell Monte Carlo "(MC)$^2$" method. Specifically, the existing MC code, based on a methodology proposed by Ghazisaeidi[16], was utilised in order to predict stable phases and phase fractions for an MPEA of choice and a given set of conditions (temperature and pressure). Using an (MC)$^2$ predicted structure, we then generated surface slabs, and performed high-throughput DFT calculations to understand the reactivity of these surfaces in aqueous environments - ahead of developing machine-learning (ML) models to predict the DFT-derived electronic energy. This was followed by a survey of surface electrochemical activity on (100) AlCrTiV surfaces with different surface features (of both the ratio and the combination of elements). We carefully investigated their selectivity and electronic response toward different species in aqueous media, including $H^+$, $O^{2-}$ and $Cl^-$. We used combined representation of elemental properties (EP)[17], smooth overlap atom position (SOAP)[18] and group and period-based coordination atom fingerprints (GP-CAF)[19] to predict the electrochemical activity: passive film formation and surface dissolution from a combination of different MPEA surfaces.

### *2.1 Bulk Calculations (MC)$^2$ Method*

The bulk structure was generated using an implementation[15] of the (MC)$^2$ algorithm[16,20,21]. The simulation was performed at T = 300 K, P = 0 Pa. Four simulation cells were initialised with 64 atoms per cell in the initial configurations of body-centred cubic (bcc), hexagonal close packed (hcp), and face-centred cubic (fcc), respectively. During the simulation, two types of moves were considered, a local flip or intra-swap. A local flip is described as randomly selecting one of the four simulation cells, then randomly selecting one atom within the chosen simulation cell, and "flipping" it from its current species type to one of the other four species types. The intra-swap consists of randomly selecting one of the four simulation cells, then randomly selecting two atoms whose positions are swapped. The algorithm has been constructed to only perform an intra-swap between two atoms of different types and should a simulation cell become 100% of one species, the intra-swap move is rejected. DFT calculations were executed to calculate simulation cell internal energy and volume changes. The acceptance probabilities, based on the Metropolis criteria[22], for the local flip and intra-swap, are given in the following formulas respectively;

$$P_{accept}^{flip} = \min\{1, \exp(-\beta\Delta H + N\Delta G_m)\} \qquad \text{(Eqn. 1)}$$

where $\Delta H$ and $\Delta G_m$ are calculated as follows:

$$\Delta H = m\sum_{i=1}^{m}(U_i' + pV_i')f_i' - \sum_{i=1}^{m}(U_i + pV_i)f_i \qquad \text{(Eqn. 2)}$$

$$\Delta G_m = m\sum_{i=1}^{m}[f_i'\ln(V_i') - f_i\ln(V_i)] + \sum_{i=1}^{m}f'^i\sum_{i=1}^{m}X_j'^i\ln(X_j'^i) - \sum_{i=1}^{m}f^i\sum_{i=1}^{m}X_j^i\ln(X_j^i) \qquad \text{(Eqn. 3)}$$

Here, $\beta = 1/k_BT$, where $k_B$ is the Boltzmann constant, $N$ is the sum of all the particles across all simulation cells, m is the total number of simulation cells, $U_i$ is the energy of simulation cell $i$, $V_i$ is the volume of simulation cell $i$, $p$ is the pressure (set to 0 Pa) and $f_i$ is the molar fraction of simulation cell $i$. Lastly, where $n_j^i$ is the number of species $i$ in simulation cell $j$, $X_j^i = n_j^i/\sum_{k=1}^{5}n_j^k$, and represents the atomic concentration of species $i$ in simulation cell $j$. The primed coordinates indicate post-flipped values, while un-primed are pre-flipped values. The updated phase fractions were obtained by using the Lever rule to enforce mass conservation.

DFT calculations were performed using the Projector Augmented Wave (PAW) method as implemented by the Vienna Ab initio Simulation Package (VASP)[23,24]. The calculations were completed with a plane-wave cut-off energy of 450 eV and a 2×2×2 Monkhorst-Pack k-point mesh[25]. DFT calculations performed on the simulation cells allowed for changes in the volume and atomic positions (through setting ISIF = 3). The electronic self-consistent calculation was converged to 1×10$^{-6}$ eV and ionic relaxation steps were performed using the conjugate-gradient method (IBRION = 2) and continued until the total force on each atom dropped below a tolerance of 1×10$^{-2}$ eV/Å. The generalised gradient approximation (GGA) was used for the exchange correlation functionals as parameterized by Perdew-Burke and Ernzerhof (PBE)[26]. The PAW pseudopotentials were used with the valence electron configurations $3s^23p^1$, $3d^34s^1$, $3d^54s^1$, and $3p^63d^44s^1$ for Al, Ti, Cr and V, respectively.

## 2.2 Surface DFT Calculation Method

All surface DFT calculations were performed using the VASP code[23]; and core electrons in calculations treated using both the the PAW method and the GGA-PBE exchange correlation functionals[24,26]. A plane-wave cut-off energy of 500 eV and a 3×3×1 Monkhorst-Pack k-point mesh[25]. The electronic self-consistent calculation was converged to 1×10$^{-5}$ eV and ionic relaxation steps were performed using the conjugate-gradient method (IBRION = 2) and continued until the total force on each atom dropped below a tolerance of 1×10$^{-2}$ eV/Å. A slab method with 20 Å vacuum thickness in z-direction was used to model (2×1) (001) surface and different slab thickness of 6, 7 and 8 layers were considered. During surface relaxation, the four uppermost layers were allowed to relax, while the others were fixed to their bulk coordinates, inducing a surface-bulk condition. **Figure 2** shows the surface of AlCrTiV MPEA which is used as the model system explored in this study.

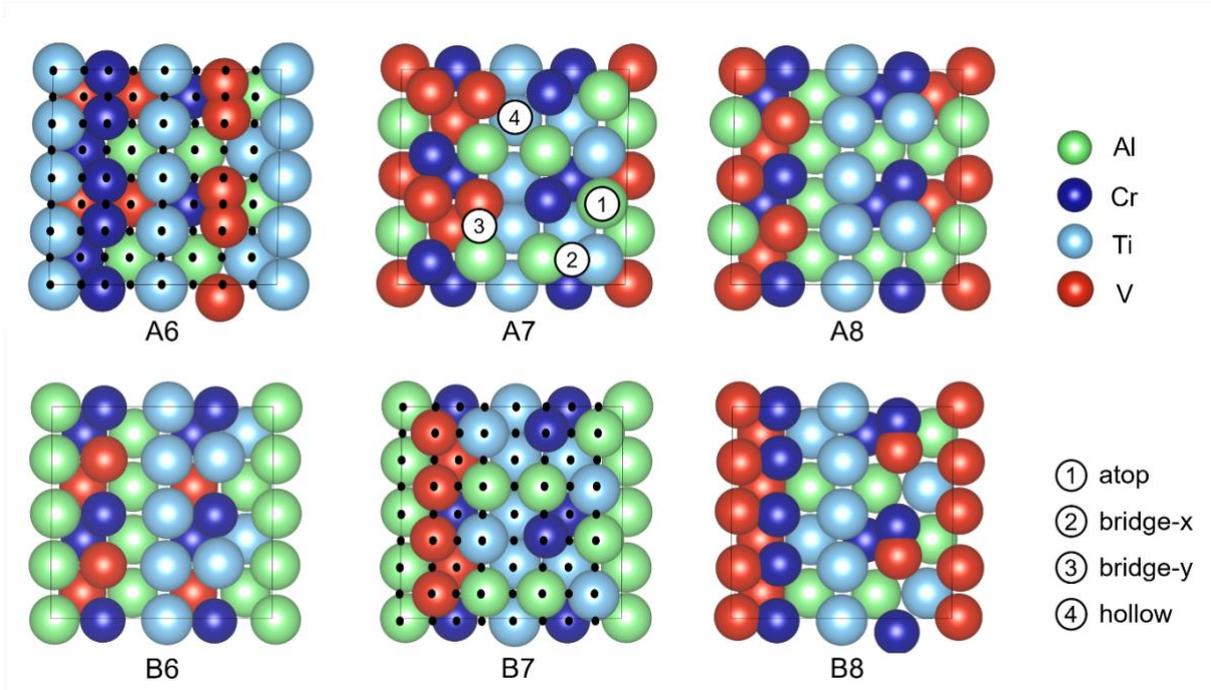

**Figure 2.** *Six different (001) surfaces of AlCrTiV bcc phase as obtained from (MC)$^2$ optimisation with four identified adsorption sites used for high-throughput DFT calculations of surface reactivity.*

Vacancy calculations were performed by removing one atom in the topmost layer on each slab. Whereas, for surface adsorption calculations, the adsorbates studied are *H, *O and *Cl, in their dissociated forms[27,28]. The adsorption study only considered single species adsorption (no co-adsorption on the surface) in each DFT calculation, in which each species on the adsorption sites was fixed in their planar (x and y) direction and only allowed relax in z-direction to find their lowest energy state. A less strict convergence criteria with total force tolerance of 5×10$^{-2}$ eV/Å was used to perform the scanning of surface reactivity/selectivity.

The adsorption energy is calculated using the following formula:

$$\Delta E_{ads} = E_{ads-slab} - E_{slab} - E_{adsorbate} \quad (\text{Eqn. 5})$$

where $E_{ads-slab}$ is the energy of adsorbate/slab complexes, $E_{slab}$ is the energy of a pure slab and $E_{adsorbate}$ is the energy of the adsorbate. Here is $E_{adsorbate}$ of H, O and Cl are defined as ½ E(H$_2$), E(H$_2$O) − E(H$_2$) and E(HCl) − ½ E(H$_2$), respectively. The influence of applied potential (U vs. SHE) and pH on ΔG can be considered implicitly[27,29], using the following mechanism:

* + H$_2$O + HCl + H$^+$ + e$^-$     (Eqn. 6)
*H + H$_2$O + HCl     (Eqn. 7)
*O + HCl + 3H$^+$ + 3e$^-$     (Eqn. 8)
*Cl + H$_2$O + 2H$^+$ + 2e$^-$     (Eqn. 9)

To identify the contribution of each element to the final adsorption energy, the following formula was fit to inference each element's contribution:

$$m_{Ti}E_{Ti} + m_V E_V + m_{Al}E_{Al} + m_{Cr}E_{Cr} = \Delta E \quad (\text{Eqn. 10})$$

where $m_{Ti}, m_v, m_{Al}, m_{Cr}$ are the total sum scaling normalised coordination number, which are calculated according to:

$$m_{Ele} = {n_{Ele}}/{(n_{Ti} + n_V + n_{Al} + n_{Cr})} \text{ for } Ele \text{ in \{Ti, V, Al, Cr\}} \quad (\text{Eqn.11})$$

where $n_{Ti}$, $n_v$, $n_{Al}$, $n_{Cr}$ is the number of coordinated Ti, V, Al, and Cr elements, respectively. $\Delta E$ is the DFT calculated adsorption energy of each mono atom adsorbate. $E_{Ti}$, $E_v$, $E_{Al}$, $E_{Cr}$ are the energy values that are fit using the least-squares method.

It is noted that herin, the vacancy energy is calculated by the following formula:
$$E_{Vacancy} = E_{SlabwV} - E_{SlabwoV} - E_{VE} \qquad (Eqn.12)$$
where $E_{SlabwV}$ is the energy of slab with vacancy, $E_{SlabwoV}$ is the energy of slab without vacancy and $E_{VE}$ is the reference energy of element.

When comparing the vacancy energies of different elements, the $E_{VE}$ plays a crucial role. In the present work, we have compared three differently defined approached to $E_{VE}$, namely: (1) the energy per atom of different elements in their pure crystal structure at room temperature; (2) the energy obtained by fitting the function $\sum_{i \in \{Ti,V,Al,Cr\}} n_i E_i = E_{total}$ for all vacancy and pure slabs in this study; and (3) the energy per atom in the high entropy alloy primary crystal, which serves as a universal correction value. **Figure 3** reveals the vacancy energy using different vacancy element reference energy. It was determined that V generally has the lowest vacancy energy using both the pure crystal structure energy and fitting energy, indicating that V is the element that could most easily form a vacancy. When comparing the magnitude of all three methods, the crystal structure atomic energy has the vacancy energy ranges in 0.0-2.5 eV, in comparison to -3.0 – 4.5 eV using fitted atomic energy (2nd method) and -3.0-2 eV using HEA average atomic energy (3rd method). However, for the AlCrTiV MPEA, empirical work by Choudhary[12,13] conducted atomic emission spectro-electrochemistry (AESEC)[30] dissolution profiling at the open circuit potential, revealing alloy dissolution kinetics in the following order: Al, V, Cr/Ti. In other words, the highest rate of dissolution in a 0.6 M NaCl solution was observed for Al, and the lowest rate was for the dissolution of Cr and Ti[13]. Therefore, the calculation using approach (3), is mechanistically the most appropriate in this work. This method provides a similar trend with Al as the most energetically favourable element to create a surface vacancy (**Figure 3**).

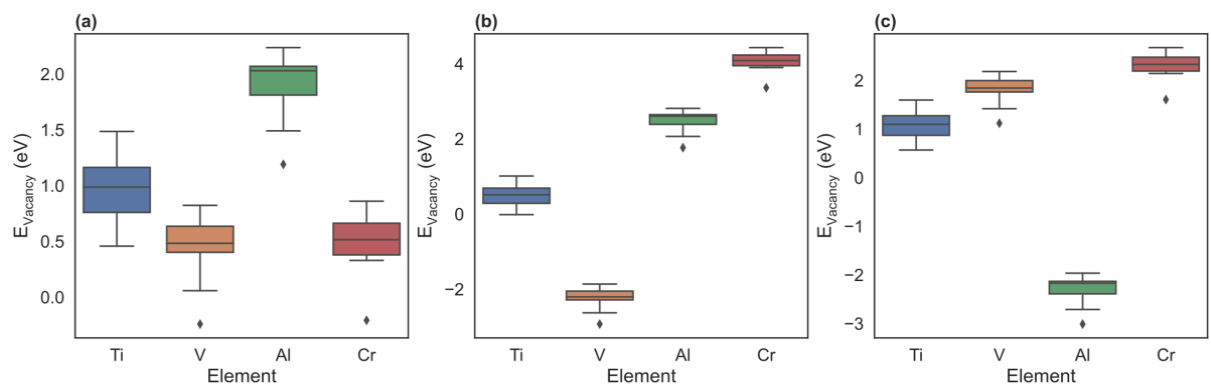

**Figure 3.** *The DFT calculated vacancy energy determined using a different vacancy reference energy: (a) the energy per atom of vacancy element in its pure crystal structure at room temperature; (b) the energy obtained by fitting the function $\sum_{i \in \{Ti,V,Al,Cr\}} n_i E_i = E_{total}$ for all slabs in this study; and (c) the energy per atom in the MPEA primary crystal, which serves as a universal correction value.*

## *2.3 ML for Surface Activity/Selectivity Prediction*

In this study, we employed Kernel Ridge Regression (KRR)[31], a fast and accurate regression algorithm[17,19,32], as the ML algorithm utilised since it demonstrated high accuracy with a small number of samples (<1000). The radial basis function (rbf) kernel was used to transform the representations of $i^{th}$ and $j^{th}$ samples into a kernel matrix $K$. The element in row $i$ and column $j$ of $K$ is calculated using the following formula:

$$K(i,j) = exp\left(-(x_i - x_j)^2 / 2\gamma^2\right) \quad \text{(Eqn.13)}$$

where $x_i$ and $x_j$ represent the representations of the $i^{th}$ and $j^{th}$ samples, $\gamma$ is a length scale parameter.

The prediction form of KRR is as follow:
$$y_{pred} = Kw \quad \text{(Eqn.14)}$$
where $w$ is a weight matrix, K is the kernel matrix.

The loss function of KRR is a quadratic function given by the following formula:
$$loss = \|y_{true} - Kw\|_2^2 + \alpha/2 \, w^T Kw \quad \text{(Eqn.15)}$$
where $y_{true}$ is the labels of the training set, $\alpha$ is a L$_2$ regularization term. A closed form solution of the loss function can be derived, as follow:
$$w = (K + I\alpha)^{-1} y \quad \text{(Eqn.16)}$$

To find the optimal hyperparameters γ and α, a grid search technique was employed. We searched for the optimal values using a base-2 logarithmic grid from 0.25 to 4096 for the kernel width and a base-10 logarithmic grid from $10e^{-7}$ to $10e^{-7}$ for the L$_2$ regularisation term.

## 3. RESULTS AND DISCUSSION

### *3.1 First-principles investigation of bulk structure*

In this study – as already ascertained from the Approach section - exploration of the AlCrTiV MPEA was carried out, owing to the alloy being known to exhibit the formation of only a single (bcc) phase at room temperature; along with a reported record regarding the contribution of each element to electrochemical reactions. The results of our attempt at the prediction of dissolution characteristics for AlCrTiV, using the workflow proposed herein; follows below, including correlation with experimental data. The $(MC)^2$ methodology accurately predicted the formation of a single bcc phase for AlCrTiV with an equi-atomic configuration and a lattice constant of 3.05 Å, with two supercells made are shown in **Figures 4(a-b)**. The final structure matched experimental data, well, with whereby the empirically determined lattice constant is 3.075 Å[10]. The empirical and simulated x-ray diffraction patterns for AlCrTiV crystal are shown in **Figures 4(e-f)**.

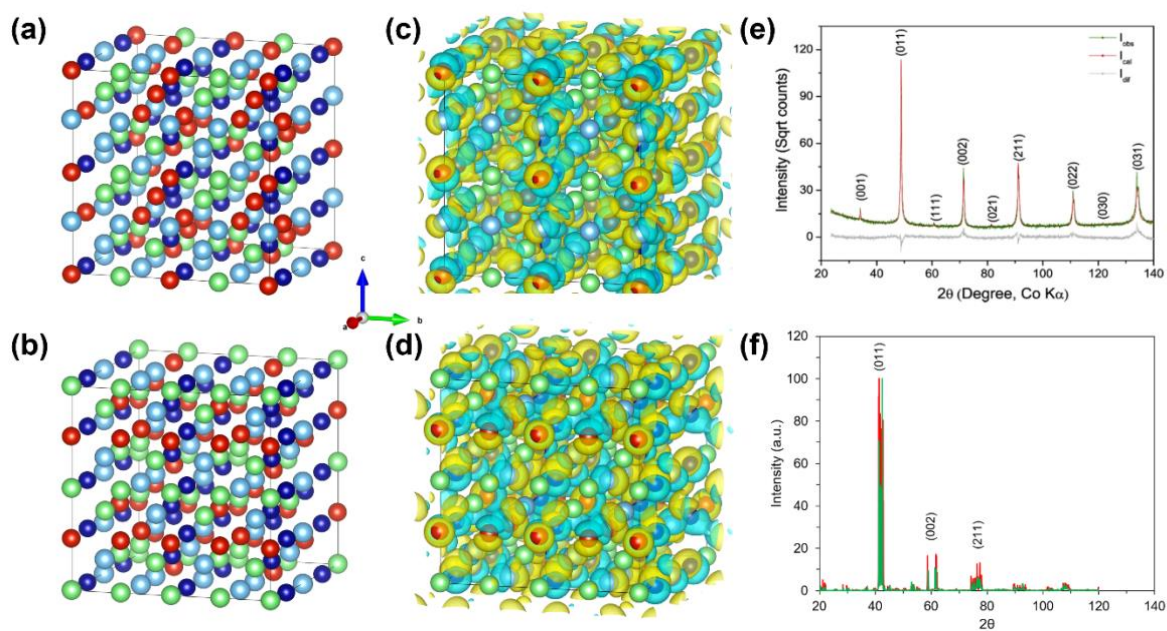

**Figure 4.** *(a, b) Two final bulk MPEA supercell structures consist of 128 atoms as generated from $(MC)^2$ calculations along with (c, d) their charge distribution. Green, dark-blue, light-blue and red circles represent Al, Cr, Ti and V atoms, respectively. Turquoise and yellow color isosurfaces represent the gain and loss of electrons, respectively. (e) X-ray diffraction pattern from a powdered sample of the as-cast AlCrTiV alloy. Reproduced with permission from Qiu et al., Acta Materialia 123, 115-124 (2017)[10]. Copyright 2017 Elsevier. (f) Simulated x-ray diffraction patterns from two final bulk structures predicted using $(MC)^2$ algorithm.*

### *3.2 First-principles investigation of surface electrochemistry*

Using the bulk AlCrTiV alloy structure predicted by $(MC)^2$, surface slabs were generated and used to build a surface reactivity map based upon the chemical interactions between different species (i.e., $H^+$, $Cl^-$, and $O^{2-}$) and the alloy surface. On the basis of empirical data being available for the electrochemical activity of AlCrTiV (which is from a bulk polycrystal), we selected the (001) surface for the adsorption model – as it provided 32 possible adsorption sites (i.e., atop, bridge-x, bridge-y, and hollow), along with eight vacancy sites. Furthermore, we also employed an extended supercell in order to minimise the lateral interaction between adsorbed species. Despite the species being charged in the electrolyte, we have only considered the adsorption of neutral species on the surface in our DFT calculations (*H, *Cl and *O); noting that the accounting for charge/electrons may be approached using the standard hydrogen electrode method[33]. To maximize atomic variation and surface structure, six different surface slabs were created for the adsorption study and a vacancy was created on the topmost layer to simulate a dissolution event at the surface. From DFT calculations, we successfully established the surface activity/selectivity that is represented using the energy map, as shown in **Figure 5**.

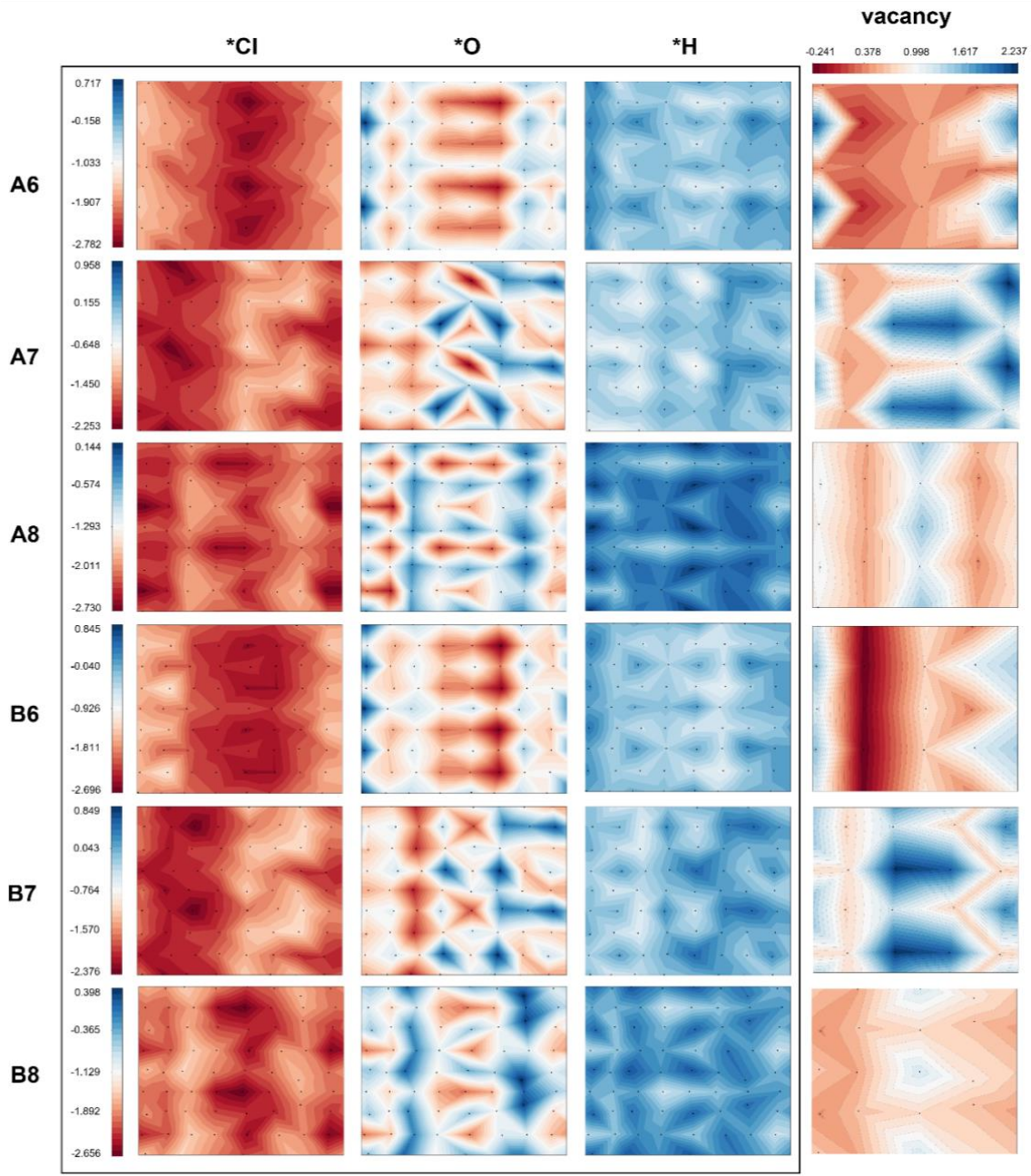

**Figure 5.** *The surface activity/selectivity maps at $U=0$ $V_{SHE}$ and $pH=0$ for six different slabs studied here (as depicted in Figure 2), showing the interactions of surface atoms following the adsorptions of $Cl^-$, $O^{2-}$, $H^+$ and the formation of vacancy.*

The results presented in **Figure 5** provide information on the early-stage electrochemical surface activity of AlCrTiV in the presence of different adsorbates. From the calculated adsorption and vacancy energy, the local electrochemical activity on the surface is observed, in which the formation of cathodic and anodic sites can be distinguished by their selective interaction with adsorbates/vacancy. Red-coloured regions indicate strong interaction between alloy-environment, whereas blue-coloured regions indicate weak interaction between alloy-environment. Strong H-bonding surface denotes the preference of surface regions to act as cathodes, whereas strong O-bonding, Cl-bonding as well as vacancy formation denote the preference of surface regions to act as anodes. Furthermore, we can understand the dissolution mechanisms of the alloys, whether it undergoes direct or indirect dissolution via salt-formation (e.g., by the formation of metal oxide and/or metal chloride). This mapping enables surface activity/selectivity to be inspected 'virtually', in a manner that is usually only obtained using complex local electrochemical characterisation techniques such as scanning tunnelling microscopy (STM) [34,35], whilst alternatively, the scanning vibrating electrode technique (SVET)[36] or scanning electrochemical cell microscopy (SECM)[37,38] provide excellent spatial resolution at an order of magnitude lower length scale. It is noted that the length scales of interrogation from the calculations herein, may also be mechanistically advantageous on the basis of prospects for nano-engineering alloy structures.

The selection of adsorbates explored was based on their critical role during surface electrochemistry (namely, alloy corrosion), such as participating in the hydrogen evolution reaction, oxide formation and thickening, surface dissolution, and pitting. From calculated surface maps, a different reactivity between may distinguish between cathodic and anodic sites on the surface. While previous studies on conventional (non MPEA) alloys have only highlighted the competitive adsorption of such species on the alloy surface as the coverage function[27,39], in the present there is a deliberate focus on the interaction between those species with the different elements on the MPEA's surface – allowing identification and analysis of their roles on the surface electrochemistry.

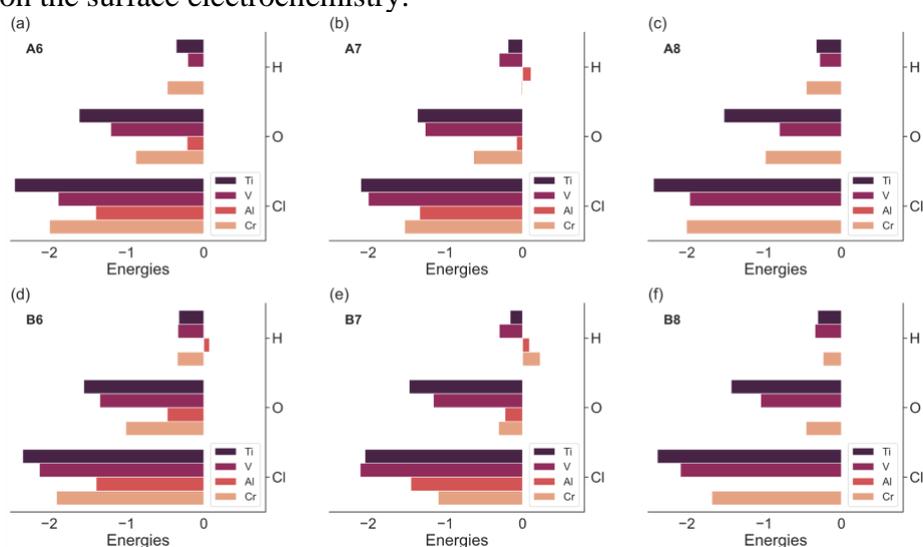

**Figure 6.** *The least-square fitted $E_{Ti}$, $E_v$, $E_{Al}$, $E_{Cr}$ on each slab at U=0 $V_{SHE}$ and pH=0.*

**Figure 6** shows the least-square fitted energy values for atop site adsorbates on each slab (where only the atop site is explored in this study, data shown in **Figure S1**). It was revealed that *O and *Cl prefer to stay close to Ti on all surfaces, except on B7 where *Cl also prefer to stay close to V. Meanwhile *H prefers to stay close to Cr on A6, A8 and B6, to V on A7, B7 and B8. From the small sampling performed, we can understand the role of each element as anodic and cathodic sites during electrochemical reactions. Ti is expected to be oxidised and formed oxide layers, the anodic site, followed by V, Cr and Al. The same order of elements for another anodic site applies for competing oxidation/dissolution via reaction with Cl⁻. The elemental order for cathodic site as predicted by hydrogen evolution reaction is varied from the systems studied here, in which Ti, V and Cr exhibits the capability to facilitate *H adsorption with negative energies, excluding Al with positive energy. This suggests that Al is more dominant and active towards any oxidation/anodic reaction, which is consistent with experiments[12,13]. Additionally, the role of Ti on the formation of protective oxide layer, including the enrichment of Ti, can be explained from its strong interaction with $O^{2-}$ and $Cl^-$. The adsorption energy difference on different slabs may be used as the preference indicator for each slab tendency as being more or less anodic/cathodic, as described by the adsorption of *O and *Cl or *H, respectively.

The inspection of **Figure 6** reveals that it is difficult to clearly distinguish the contribution of each element, due to the complexity of alloy systems (i.e. in addition to chemical complexity, there is also a variation of adsorption sites, which includes the bridge and hollow sites that require the consideration of multiple atoms) – as well as the limitation of slab samples investigated in this study. We acknowledge that the electrochemical activity of each element is unique and not solely defined by its own intrinsic features. Neighbouring atoms and a coordination network will contribute to electrochemical activity, as understood from the unique charge distribution of each atomic species in MPEA structure in **Figures 4(c-d)**. Therefore, we will refrain from quantitatively focusing on feature rank and selection during the present work. The future objective development of new MPEAs with controlled electrochemical activity will require the theoretical analysis of many samples, which can also become an obstacle even in theoretical-based studies. Instead, we are more interested in the ongoing generation of datasets for training ML models to predict the electronic energy of certain surface structures. This approach will be beneficial for the generation and analysis of large surface datasets, to better understand the contribution of individual elements on surface electrochemistry – which is particularly important for MPEAs.

From the experimental work to date studying AlCrTiV, it is understood that 'dynamic-passivity' is a critical factor contributing to corrosion resistance of AlCrTiV; including the surface film composition, thickness and electronic properties. Various analysis tools have confirmed that AlCrTiV exhibits excellent electrochemical stability, which is attributed to the ability to form passive layer comprising of mixed metallic oxides: $Al_2O_3$, $TiO_2$, $V_2O_3$ and $Cr_2O_3$. The experimental techniques used to study the passive film characteristics and electrochemical activity, include high resolution X-ray photoelectron spectroscopy (XPS) and atomic emission spectroelectrochemistry. Using AESEC, Choudhary *et al.*[12,13] quantitatively recorded the dissolution and oxidation of AlCrTiV on a 'per element' basis. It was revealed that elemental Al was readily observed to preferentially dissolve (although the alloy has stoichiometrically equal atomic proportion of all constituent elements), as soon as the sample was exposed to the electrolyte, as shown in **Figure 7(a)**.

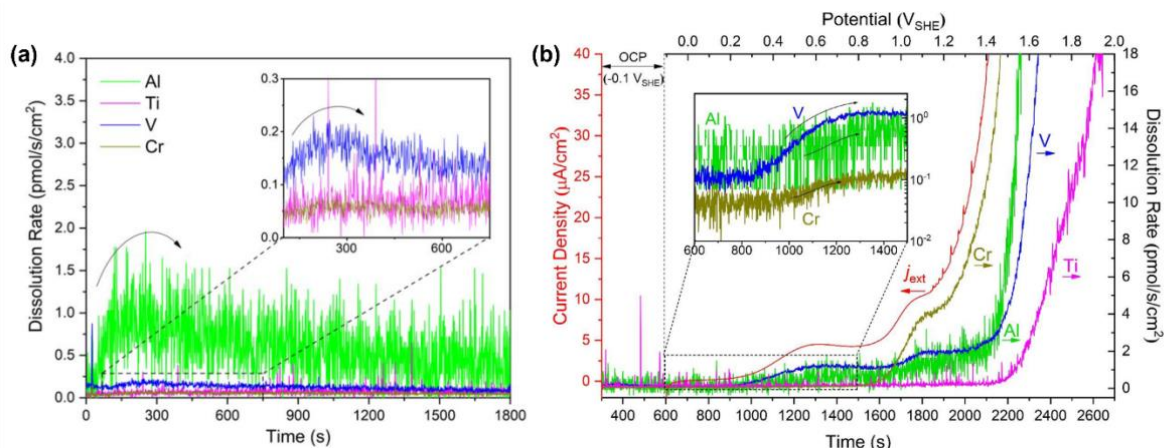

**Figure 7.** *(a) AESEC dissolution profile at OCP for AlCrTiV in quiescent 0.6 M NaCl. Reproduced under Creative Commons CC-BY license from Choudhary et al., J. Electrochem. Soc. 168, 051506 (2021)[13]. (b) AESEC polarization profile for AlCrTiV in quiescent 0.6 M NaCl. Reproduced with permission from Choudhary et al., Electrochim. Acta 362, 137104 (2020)[12]. Copyright 2020 Elsevier.*

It is also revealed in **Figure 7(b)**, that in the presence of an applied potentiodynamic polarisation, that trans-passive dissolution of V and Cr occurs, accompanied by the enrichment of Ti and its oxidation to form $TiO_2$ on the surface[12,13]. Despite the high dissolution rate of elemental Al, results from detailed XPS analysis revealed that $Al_2O_3$ was the major oxide present on the surface, followed by $TiO_2$ and then similar amounts of $V_2O_3$ and $Cr_2O_3$ [13]. The absence of mixed oxides here is also consistent with the findings of the computational study by Samin on the thermodynamics of Niobium-Titanium alloy oxidation[40]. In that study, it was determined that $TiO_2$ was found to be the most stable oxide for most temperature-pressure combinations and mixed oxides were never thermodynamically favourable. Key findings from such aforementioned studies regarding corrosion resistant alloys, are that dissolution/oxidation are non-stoichiometric, that mixed-oxides are not present, and that the surface oxides also include a proportion of unoxidised ($M^0$) metal. The causality between such characteristics and performance however remains under ongoing investigation.

*3.3 Machine-learning prediction of surface electrochemistry*

In the workflow outlined for this study, the utilisation of ML was then adopted for predicting adsorption and vacancy energies. In a previous study, Batchelor *et al.* employed a coordination-based representation and linear regression to predict adsorption energies of O and OH on high entropy alloys[41], resulting in an RMSE of less than 0.1 eV. Nonetheless, that method was developed to solely rely on surface coordination and hence disregards the identity of the adsorbed species; therefore, requiring separate models for each adsorbate. Li et al. proposed an alternative approach that integrates both adsorbed species and surface slab information[17], enabling the prediction of "cross-adsorbate" and "cross-slab" adsorption energies. That representation, however, is limited to the most stable site and does not consider specific adsorption sites or lateral interactions between adsorbates.

The present study enhances the combined representation by incorporating an additional representation for the adsorption site, enabling "cross-site" prediction. The improved representation is comprised of EP to represent the single-atom adsorbate (*O, *H, or *Cl in this study), SOAP to represent the high entropy alloy slabs[18], and GP-CAF to represent the adsorption site[19] (representation employed herein shown schematically in **Figure 8**).

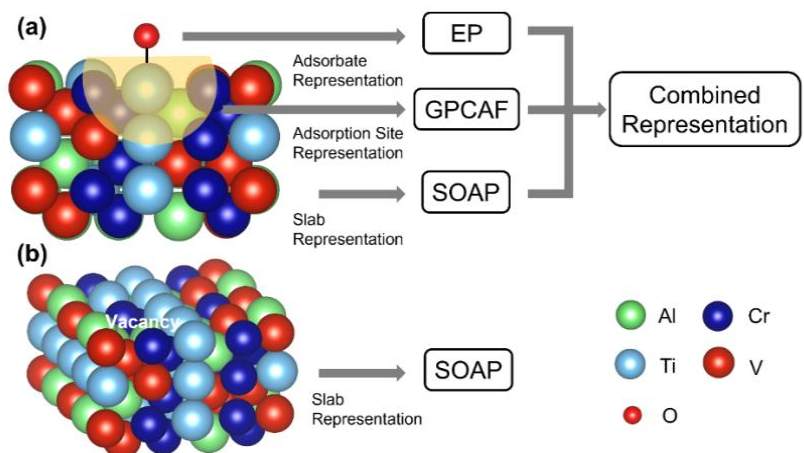

**Figure 8.** *Schematic of the combined representation used in this study. Elemental properties (EP) is used as adsorbate representation, group and period-based coordination atom fingerprints (GP-CAF) as adsorption site representation and smooth overlap atom position (SOAP) as slab representation.*

**Figures 9(a-c)** compare the ML-predicted adsorption energies and DFT calculated values using 5-fold cross validation test, leave-one-slab-out test (LOSO), and LOSO with 20% samples from the test slab added to the training set. The 5-fold cross validation test results show a strong correlation between the ML predictions and DFT calculated values, with a mean average error (MAE) of 0.197 eV, highlighting the effectiveness of our model. In practice, it is desirable for ML methods to have good "cross-slab" prediction capabilities, as predicting the adsorption activity on new MPEAs is a key goal of these calculations. The LOSO test was designed to evaluate the model's "cross-slab" prediction abilities, as seen in **Figure 9b**. This test resulted in a higher MAE compared to 5-fold cross validation test and worse predictions for outliers, such as the B6 slab, where the ML model systematically underestimated the adsorption energies. However, our model accurately distinguished high and low adsorption energies on the test slab, indicating its ability to correctly identify the most stable adsorption site. The accuracy of LOSO test can be improved by adding 20% samples from the test slab to the training set, as shown in **Figure 9c**, and would likely to be improved with more slabs in the training set as there are only 6 slabs in this study.

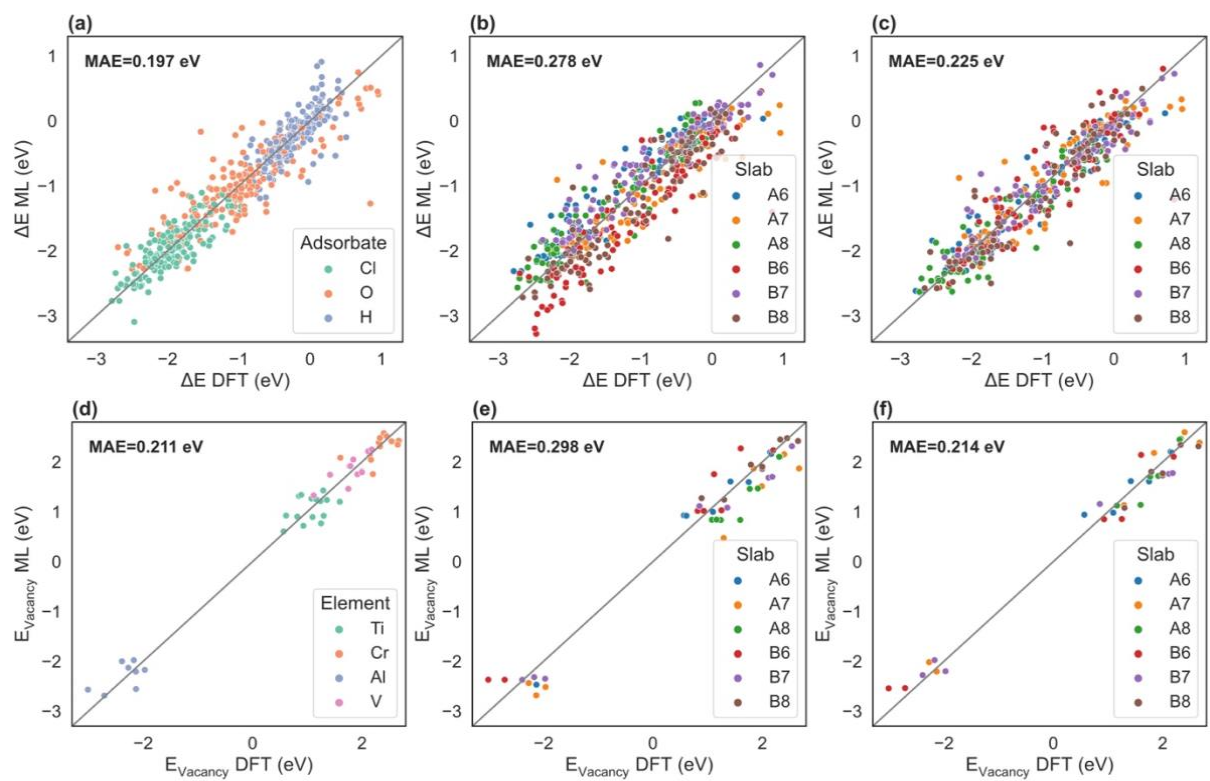

**Figure 9.** *ML predicted adsorption energies against DFT calculated adsorption energies using (a) 5-fold cross validation (b) leave-one-slab-out test, and (c) leave-one-slab-out with 20% samples from the test slab added to the training set. ML predicted vacancy energies against DFT calculated vacancy energies using (d) 5-fold cross validation (e) leave-one-slab-out test, and (f) leave-one-slab-out with 25% samples from the test slab added to the training set.*

Besides testing the performance of the ML model in predicting adsorption energies, we also evaluated its capability in predicting vacancy energies. **Figures 9(d-f)** shows a comparison between the ML-predicted adsorption energies and DFT calculated values, using 5-fold cross validation test, LOSO, and LOSO with 25% samples added from the test slab to the training set (25% means 2 samples on test slab considering there are 8 vacancy sites on each slab). Similar to the adsorption energy prediction, the model performed best using 5-fold cross validation test, as the training and test set have similar distributions. Also, LOSO test was significantly worse than 5-fold cross validation testing and could be improved by adding 25% samples from the test slab to the training set.

### 3.4. Integrated computational approaches as a surface electrochemistry simulator

We have shown the capability of ML to replace the use of high-throughput DFT for providing data on the reactivity of the surface models of an alloy. Such results demonstrate the effectiveness of ML modes in accurately predicting adsorption of different species and vacancy energies on MPEA. The key findings from the work herein include:

(i) the application of the (MC)$^2$ approach was capable of providing a platform for subsequent DFT and ML approaches – that are relevant to the determination of electrochemical properties – where the prediction of electrochemical properties has notionally been historically challenging to model,

(ii) the workflow herein provides a visual, spatial map of electrochemical properties from calculations entirely *in-silico*,

(iii) all elements in AlCrTiV exhibit unique properties due to different charge distributions in MPEA compared to their pristine crystal structure, in which Al exhibits spontaneous formation of vacancy in AlCrTiV alloy that is well validated by independent empirical data[12,13] with the observation of high Al dissolution rate,
(iv) the simulations were also able to ascertain that for AlCrTiV, Ti has the most important contribution on the formation of protective oxide species such as $TiO_2$,
(v) the use of DFT and ML to model and predict AlCrTiV surface activity/selectivity by considering different species, which are critical for reduction oxidation reactions, enable the investigations of nano-scale electrochemical reactions on a complex surface.

Whilst we believe the above findings, and the pathway to obtaining such computational results regarding surface electrochemistry are significant, it is also prudent to identify some of the physical limitations, and required future work, in order to gain the most benefit from such approaches when applied to rationalising MPEA behaviour. Some points that remain open and will be the focus of future studies include:

- As also noted in this study, the corrosion performance of MPEAs is closely linked to the nature of the surface films. The modelling approach herein is in part, a proxy to studying surface films on the basis that the modelling approach deals with a 'snapshot' in time for an alloy surface (whereas surface films develop dynamically). The model is indeed a major simplification of the MPEA-electrolyte interface, as it presently stands – however this is typical for any early model that may evolve in complexity.
- The numbers of sample for MPEA's surface, AlCrTiV, studied here using DFT are small. Large datasets are necessary to make any conclusive statement regarding elemental's role on the electrochemical reactions *in-silico*.
- Additional features, such as different surface orientations, explicit modelling of electrolyte system including cations and anions[27,42], explicit treatment of temperature, pH and applied potential[27,29], will be beneficial for a more accurate prediction of MPEA's electrochemical activity as well as to understand multiple competing reactions on the surface. A detailed investigation on competing adsorptions in changing pH and applied potential[27,39] either implicitly (as described using Eqns. 6 – 9) or explicitly, can give insights on the subsequent pathway (i.e., atomic dissolution, hydrogen evolution reaction, oxide formation).
- Large datasets are also necessary to improve the performance of our ML prediction algorithm, which will be the key of such an *on-the-fly* approach.

The approach outlined in the study herein, whilst applied to only a single empirical benchmark, has provided a computation workflow that may be utilised as a virtual electrochemical characterisation tool. Such a tool has the capacity to predict and therefore estimate the electrochemical properties of MPEAs *in-silico*, by considering composition, crystallographic orientation, and number of samplings. As a result, the computation workflow herein introduces a cheaper and faster approach to garner an insight into electrochemical properties of new MPEAs with tuned composition and phase, designed entirely from simulations; with MPEAs revealing promising electrochemical performance laboratory verified accordingly.

## 4. CONCLUSIONS

Herein, a methodology has been presented for generating realistic MPEA structures for AlCrTiV, using implementation of the $(MC)^2$ algorithm for the investigation of electrochemical activity and development of activity/selectivity maps. Furthermore, the work herein reports on the development and utilisation of an ML method which capable of predicting surface electrochemical activity *via* adsorption and vacancy energies. Independent detailed experiments were used to correlate and verify the electrochemical properties of the AlCrTiV MPEA, derived from the simulations herein - in which Ti has the most important contribution on the formation of protective oxide species such as $TiO_2$.

The combined $(MC)^2$ and DFT/ML approach presented in this work is a potential candidate for intelligently exploring vast numbers of elemental combinations (MPEA compositions); which is both critical and necessary in providing an *in-silico* insight for rationalising structure-electrochemistry relationships. One of the key features is the identification of active sites *via* the construction of an electrochemical activity/selectivity map. Such mechanistic understanding is beneficial for building an improved electrochemistry microkinetic model for MPEAs, in which reactions are non-uniform, and each reaction is sensitive to local surface features. The approach developed herein is readily applicable for the design and application to other MPEAs.

**DECLARATION OF COMPETING INTEREST**
The authors declare that they have no known competing financial interests or personal relationships that could have appeared to influence the work reported in this paper.

**CODE AND DATA AVAILABILITY**
Our implementation of $(MC)^2$ code for generating the bulk structure is available at: https://github.com/SaminGroup/Dolezal-MC2.
The DFT calculation was executed using VASP software.
The ML code for energy prediction is available upon request.
Data generated from this study is all published in this study and available upon request.

**AUTHOR CONTRIBUTIONS**
**Jodie A. Yuwono**: Conceptualisation (lead); Resources (lead); Research (lead); Writing – original draft (lead); Writing – review & editing (lead). **Xinyu Li**: Research (equal); Writing – original draft (supporting). **Tyler D. Doležal**: Research (supporting); Writing – original draft (supporting); Writing – reviewing & editing (supporting). **Adib J. Samin**: Resources (supporting); Writing – review & editing (supporting). **Javen Qinfeng Shi**: Resources (supporting); **Zhipeng Li**: Writing – review & editing (supporting). **Nick Birbilis**: Resources (lead); Writing – original draft (equal); Writing – reviewing & editing (lead).

**ACKNOWLEDGMENTS**
The authors acknowledge the high-performance computing (HPC) resources from the Department of Defense through Air Force Research Laboratory (AFRL) HPC Mustang and from National Computing Infrastructure (NCI) Australia through NCI Gadi. X.L. and J.Q.S. acknowledge the financial support from the Center for Augmented Reasoning, Australian Institute for Machine Learning. T.D and A.S. acknowledge the financial support from the Air Force Office of Scientific Research (AFOSR). Financial support from the Office of Naval Research under the contract ONR: N00014-17-1-2807 with Dr. David Shifler and Dr. Clint Novotny as program officers is gratefully acknowledged.

**Graphical abstract**

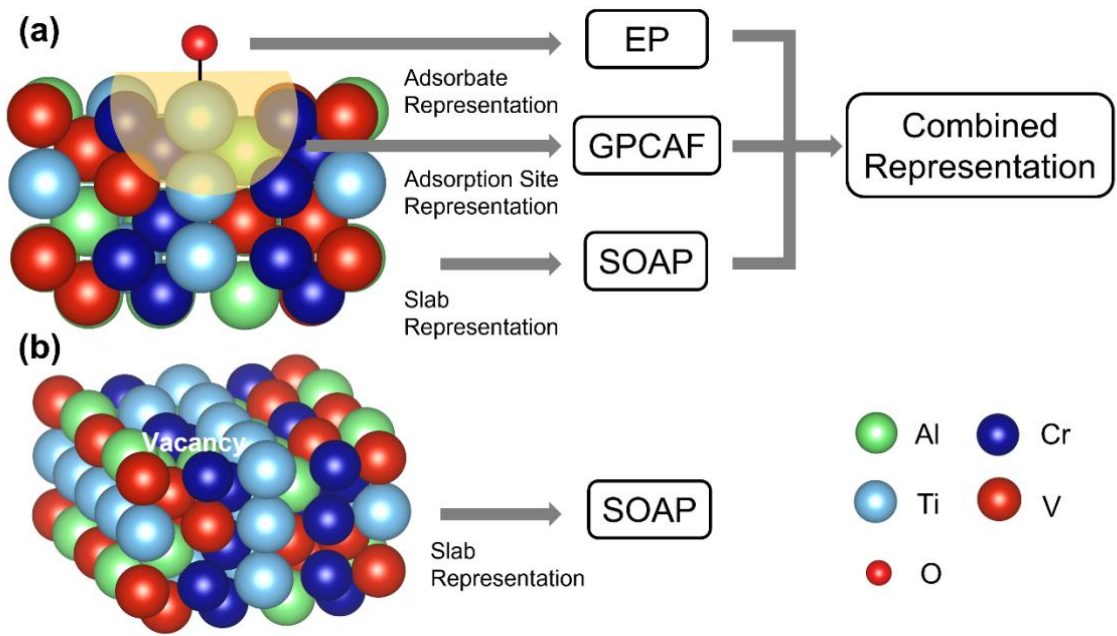

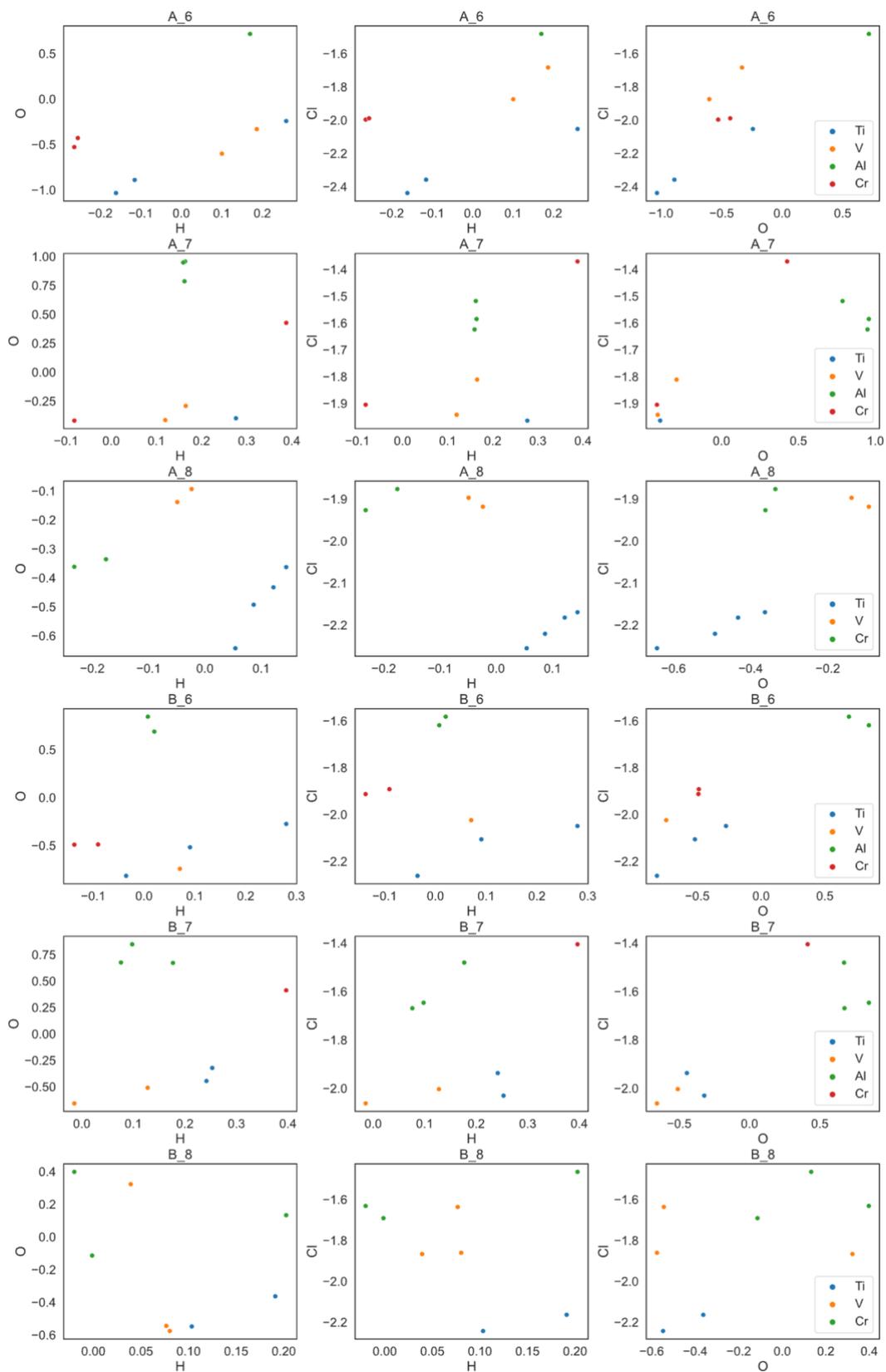

**Figure S1.** *The adsorption energies of *O, *H and *Cl on atop sites for each slabs of the AlCrTiV (001) surface.*